# Web Analytics for Security Informatics

Kristin Glass
New Mexico Institute of Mining and Technology
Socorro, NM USA
kglass@icasa.nmt.edu

Richard Colbaugh
Sandia National Laboratories
Albuquerque, NM USA
colbaugh@comcast.net

*Abstract*—An enormous volume of security-relevant information is present on the Web, for instance in the content produced each day by millions of bloggers worldwide, but discovering and making sense of these data is very challenging. This paper considers the problem of exploring and analyzing the Web to realize three fundamental objectives: 1.) security-relevant *information discovery*; 2.) target *situational awareness*, typically by making (near) real-time inferences concerning events and activities from available observations; and 3.) *predictive analysis*, to include providing early warning for crises and forming predictions regarding likely outcomes of emerging issues and contemplated interventions. The proposed approach involves collecting and integrating three types of Web data – textual, relational, and temporal – to perform assessments and generate insights that would be difficult or impossible to obtain using standard methods. We demonstrate the efficacy of the framework by summarizing a number of successful real-world deployments of the methodology.

I. INTRODUCTION

There is increasing recognition that Web data, such as webpage content, dialog threads between forum participants, and relationships in social networking sites, are important sources of information for intelligence and security applications [e.g. 1-5]. For instance, vast volumes of security-relevant information are generated each day by bloggers and other content producers worldwide, thereby providing an essentially real-time view of the opinions, activities, and trends of individuals and groups across the globe. These data may, for instance, allow early detection of emerging issues and trends in populations of interest, which could be of considerable value [e.g. 6]. For example, contentious situations are much easier to resolve if discovered in their early stages, while nascent positive sentiments and activities can often be leveraged and amplified. Of course, these security-relevant data are buried in the massive output of millions of online content generators, so that discovering and interpreting this information rapidly enough to be useful is very difficult.

Recognizing both the opportunities and challenges associated with analyzing Web data for security applications, several researchers have studied this problem in recent years [1-11]. One interesting theme is the assessment and characterization of the ways in which the Web is used by jihadists, criminals, extremists, and others of interest to security professionals [e.g. 2,3,7-9]; indeed, a key motivation for Web-based security informatics is the active online presence of these individuals and groups. Other researchers have focused on the problem of identifying, extracting, and understanding security-relevant information at Web-scale, and have developed collection and analysis tools which hold considerable promise [1,4,5,10,11]. Not surprisingly, other application domains face similar challenges when attempting to make sense of Web data. As one example, the burgeoning field of Web-based business informatics is receiving substantial attention from both researchers [12-14] and practitioners [15,16].

Despite much progress by security informatics researchers and analysts, important open problems remain. For instance, although a defining feature of Web data is the simultaneous presence of textual, relational, and temporal information, little has been done to develop methods for systematically integrating and exploiting these complementary data classes. Additionally, many of the techniques currently available are not well-suited for Web-scale analysis, which may involve tens or hundreds of millions of entities and relationships. Another concern is the fact that analytic tools intended to answer security-related questions, say related to deceptive behavior, are often borrowed from domains with quite different operational characteristics (e.g., in which intentional deception is uncommon). Perhaps most importantly, the rich temporal information present online offers the potential to perform *predictive* analysis, for example permitting emerging threats to be anticipated and mitigated proactively [17]; however, little has been done to rigorously investigate this exciting possibility.

This paper addresses the general problem of Web-based security informatics, and the specific challenges noted above, by proposing a computational framework for exploring and analyzing the Web to realize three fundamental objectives: 1.) *discovery* of security-relevant data and information (e.g., uncovering forums devoted to extremist ideologies or detecting growing unrest in populations of interest); 2.) target *situational awareness*, typically by making near real-time inferences about phenomena of interest based on indirectly-related observables; and 3.) *predictive analysis*, to include giving early warning for conflicts and crises and forming predictions concerning likely outcomes of emerging issues and contemplated interventions. The proposed approach collects and integrates three types of Web data – textual, relational, and temporal – to perform assessments, generate insights, and provide decision support that would be difficult or impossible using standard analytics. We demonstrate the efficacy of the analysis methodology through a series of real-world examples, including studies involving detection of cyber incidents in near real-time, estimation of public opinion in contentious situations, discovery of emerging topics and trends, and early warning analysis for mobilization and protest events.

## II. WEB ANALYTICS FRAMEWORK

A simple schematic of the proposed Web analytics framework is shown in Figure 1. This system is designed to overcome the challenges associated with extracting actionable intelligence from massive, networked, dynamic, and noisy Web data. The analysis framework consists of three main capabilities: 1.) data collection/preprocessing, 2.) computational analysis, and 3.) user interaction. In this section we briefly summarize each of the process elements and the way they work together, and highlight a few of the interesting aspects of the framework. Additional details regarding the system architecture and the data collection process can be found in [18,19], and further discussion of the computational analysis and user interaction components are given later in this paper.

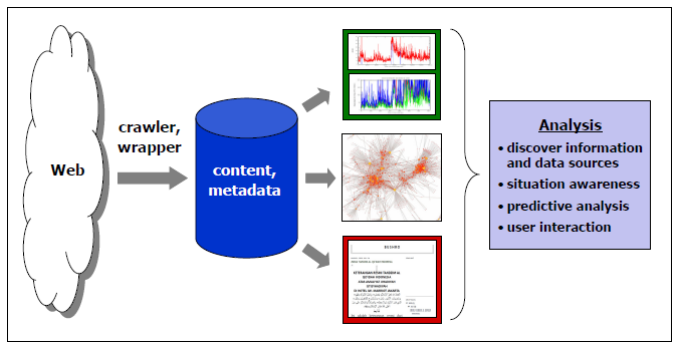

**Figure 1.** Schematic of basic Web analytics process.

The overall process flow is fairly typical. The data collection system, called CACTUS [18], first "crawls" the Web, harvesting textual data (e.g. the content of webpages), relational data (e.g. the way individuals are linked within a social networking site), and temporal data (e.g. the timestamps of posts to a blog). These data are then preprocessed to enable subsequent automated and manual analysis. The collection and pre-processing steps include both "standard" and "advanced" features. Standard capabilities available in CACTUS include: 1.) efficient and effective (e.g. threaded, throttled) crawling of web pages, blogs, and forums, based on both site-list seeding and (boolean) key word queries; 2.) fast indexing of collected data to support querying, analyst "drill-down", and advanced computational analysis; and 3.) standard wrapping and parsing tools to allow extraction of content, metadata, relational data, and temporal information. CACTUS also provides three advanced features: 1.) it can collect and process information from several specific, security-relevant social media sites (e.g. YouTube, Twitter); 2.) collected data is encoded and archived as time series of semantic graphs (i.e., graphs with a rich set of vertex/edge types and attributes [20]); and 3.) data collection is guided by our *predictability assessment framework* [21,22], thereby substantially increasing the relevance and usefulness of the harvested data.

Once data have been collected and encoded as time series of semantic graphs, computational analysis can begin. The analytic process integrates and analyzes textual, relational, and temporal data within a scalable, mathematically-rigorous, and empirically-grounded framework. As shown in Figure 1, the computational analysis phase has three main objectives: 1.) discovery of data/information sources, 2.) situational awareness, and 3.) predictive analysis. In the subsequent sections of the paper we define each of these objectives more precisely, describe the analytic approaches we've developed to accomplish them, present a sampling of tools and methods which illustrate some of the ways the approaches have been implemented, and discuss case studies in which the analytic tools have been deployed in real-world settings.

A key function of the proposed analytic framework is to enable productive and convenient analyst interaction. Thus, for instance, the CACTUS data collection and preprocessing system possesses an automated crawl summarization capability which presents to the user an intuitively accessible synopsis and characterization of a given Web crawl. Example capabilities include: 1.) organizing and displaying the results of a completed crawl according to various Web-graph layouts, 2.) summarizing the link structures and the geographic and lexical profiles of individual sites, related communities of sites, and the entire collection of crawled and collected sites, and 3.) support for analyst-directed exploration and drill-down investigations. Indeed, a central feature of both the data collection and computational analysis elements is the presence of numerous "hooks" that facilitate analyst-driven, iterative discovery and investigation.

We conclude this section by offering a sample of some key security informatics questions that we believe Web analytics can impact in important ways and which will serve as motivation for much of our development.

- Given an initial set of useful data sources in a particular domain (e.g. extremist blogs), what are additional data sources which are also relevant to the domain?

- Do there exist unknown but compelling new proliferation threats in a specified weapons domain (e.g. technologies, people)? Are there developing but presently unrecognized issues which are likely to become contentious?

- What is the sentiment of a given population concerning a particular topic? What is the associated trend?

- Who are the influential individuals and groups in a given domain? What are authoritative sources of information?

- Are the objectives of a given organization or institute what is claimed? What would be useful indicators that these objectives are changing?

- Who are the collaborators of a given individual or group? What are the plans and intentions of this collaboration?

- What are emerging topics and trends that will become important in the near term? In the longer term?

- How is a given threat situation likely to evolve? What would be the associated consequences? What are the likely effects of various intervention options?

- How can we move from a reactive to a proactive posture with respect to a given adversary or threat?

## III. DISCOVERY

This section begins by defining the classes of discovery questions of interest, then summarizes the proposed approach to conducting Web-based discovery, and finally briefly describes an implementation of one useful discovery tool: dynamic semantic graph analysis via tensor decomposition.

### A. Proposed Approach

Motivated by the kinds of questions faced by intelligence and security analysts, we are primarily interested in two classes of discovery problems: 1.) "seeded" discovery – roughly, given an initial set of examples of things of interest (e.g. events, entities), find more of these things, and 2.) "unknown-unknowns" discovery – uncover classes of activities and entities that pose a plausible threat and which analysts aren't even aware exist.

For Web-based seeded discovery, it is assumed that the examples of things of interest are Web objects, for instance webpages, blog posts, forum threads, or trends of various kinds. In what follows, we focus on applications for which the seeds are websites (e.g. blogs, forums); other possibilities are considered in [18-20]. In this case, the exemplar websites can serve as seeds from which to launch Web crawls, and the discovery problem can be formulated as one of collecting large numbers of websites and assessing the extent to which each of the new sites is "similar" to the initial seeds [18-20]. While there are numerous ways to measure the similarity of a pair of websites, we have found three simple techniques to be effective in security informatics applications.

First, one can measure similarity with respect to content. One useful approach is to begin by representing the content of each website (or subset of the site's pages) as a "bag-of-words" feature vector $x \in \Re^{|V|}$, where the entries of x are the frequencies with which the words in the vocabulary set V appear in the pages(s) (perhaps normalized in some way). Then one can employ a range of standard metrics, such as cosine similarity or Jaccard similarity [19], or more Web-specific measures [e.g. 23], to quantify the similarity of candidate site to the seed sites. Newly discovered sites which exceed some desired similarity threshold are then returned to the analyst for additional study.

The second method we use to identify websites which are "like" the seeds is to focus on (hyperlink) topological similarity. Again there are several metrics which can be adopted to quantify the similarity of a pair of websites, such as set similarity of the out-link or in-link neighborhoods of the two sites [19]. Alternatively, we have found that partitioning the webgraph obtained via crawling into *graph communities* (densely interconnected groups of websites which are only loosely connected to other groups [24]), enables effective evaluation of site similarity: similar sites often belong to identical, or closely related, communities [19]. Once topological similarity has been quantified, new sites which are "similar enough" to the seeds are returned to the analyst for further investigation.

Finally, we have found it useful to *combine* textual and relational information when measuring the similarity of a pair of websites. For instance, assuming that the content and local hyperlink topology of the two websites are modeled as semantic graphs, these graph models can be transformed into tensor models (see the discussion of the case study later in the section for a description of this transformation procedure). Once the content and topological data for each website have been encoded as tensors, their similarity can be quantified via the Frobenius tensor norm [e.g. 25]. Alternatively, the tensor models for the seeds, and for randomly selected sites, can be used to train machine learning algorithms [e.g. 26] that classify new websites as either 'similar to' or 'different from' the seeds. It is worth noting that the three techniques summarized above, although simple, have been successfully implemented for seeded discovery of new extremist websites [18,19].

The second class of discovery tasks of interest in this paper are "unknown-unknowns" problems: uncover activities, events, groups, trends, and so on which are both relevant to security analysts and presently unknown to them. It is natural to formulate such discovery problems in terms of anomaly detection. In particular, we have obtained good results by posing these questions as *structured* anomaly detection, that is, anomaly detection which leverages an understanding of the properties and characteristics of the Web [20]. As a simple but illustrative real-world example, consider the task of discovering a scientific or technological innovation which could pose a plausible threat. Because there exists a large quantity of information on the Web which has little or no scientific validity, it is necessary to develop methods capable of discovering anomalous (i.e., new) and technically-correct information. One approach to addressing this challenge is to search for scientific concepts on the Web which are treated as interesting and legitimate by one authoritative organization but not by other such organizations. The rationale for this simple scheme is that such behavior may reveal a technically sound idea which is not generally acknowledged, perhaps because it is new or it represents a threat and is thus treated as sensitive information. We have successfully applied this basic anomaly detection methodology to weapons development and proliferation problems [20].

It seems clear that in order to consistently return good answers to unknown-unknowns questions, it is necessary to combine computational methods with manual analysis in an interactive, iterative process. We have found the automated crawl summarization capability in CACTUS often provides a convenient starting point for this process, as it displays security-relevant information obtained during the crawl in an intuitively accessible format [18]. Additionally, adopting a semantic graph formalism as the underlying data model tends to increase the efficiency and effectiveness of the user interaction, for instance by supporting informative data visualizations and enabling intuitive drill-down analysis [19,20].

### B. Case Study: Tensor Analysis

In this subsection we illustrate, through a case study, an effective and efficient method for discovering and characterizing security-relevant events in large data sets associate with network dynamics. The proposed approach models the data as a three-way tensor, that is, a three-dimensional array in which each "slice" of the tensor represents network activity during a particular time interval [25]; see Figure 2 for a depiction of the basic structure. Significant events are then uncovering by examining low-rank approximations to this tensor using a procedure analogous to familiar singular value decomposition analy-sis of matrices [26]. The specific network dynamics discovery problem we study here involves the detection and localization of large Internet disruptions using data which can be collected from the Web. However, it should be clear that the proposed analytic process is applicable to a wide range of network dynamics discovery tasks. It is worth noting that tensor models employed in this case study can be productively viewed as alternative representations of semantic graphs [e.g. 20].

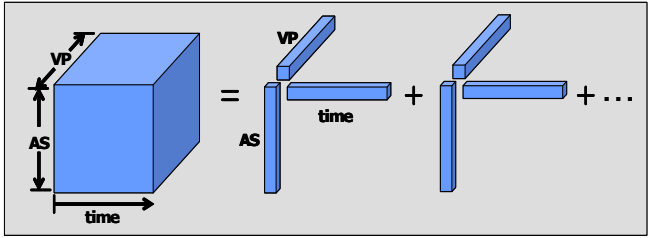

**Figure 2.** Cartoon of tensor decomposition. The original BGP update tensor is factored into a sum of rank-one tensors, each of which is an outer product of three vectors: an AS profile, an update stream vector, and a VP profile. The most dominant few rank-one tensors typically correspond to "large" events.

There is substantial motivation to develop a method for discovering large Internet disruptions and locating the "source" of each disruption, that is, identifying the network elements whose perturbation is the "root cause" of the event. But detecting and locating large, disruptive Internet events is a challenging discovery problem for at least three reasons. First, the Internet is a complex dynamical system which is very large (e.g., consisting of over 25,000 autonomous systems (AS) and millions of hosts) and about which there is significant uncertainty (e.g., because many operational characteristics are considered proprietary). Second, data sets associated with Internet behavior are massive, noisy, and give only an incomplete picture of events of interest. Finally, the events themselves are difficult to understand and characterize, for example because of their immense number, wide heterogeneity, significant area of impact, and evolving nature.

We present in [27] a new approach to detecting and locating large Internet events which addresses these challenges. The proposed methodology consists of three steps: data collection, event detection, and logical localization. Data collection involves acquiring Border Gateway Protocol (BGP) route update data by crawling the RIPE website [28] and assembling these data into an analytically-tractable format. Note that RIPE maintains BGP peering sessions with several Internet routers and logs the updates they receive, so that these data provide views of Internet activity from numerous vantage points (VP).

Event detection is accomplished via tensor decomposition-based analysis of BGP time series. We model the Internet as a graph at the AS level, $G_{AS} = (V_{AS}, E_{AS})$, where each vertex $v \in V_{AS}$ is an AS (collection of connected routing prefixes that presents a common routing policy to the Internet) and an edge $e = (v, v') \in E_{AS}$ reflects the fact that routing policy allows traffic to be sent directly from v to v'. The dynamics of $G_{AS}$ are governed by BGP, the protocol by which neighboring routers keep each other apprised of changes in prefix reachability.

BP update data is used to assemble an update tensor $\underline{X}$ by first specifying a sampling period $\Delta t$ (e.g., $\Delta t$=30s), an observation interval $T \cdot \Delta t$ (e.g., T=2880, which with $\Delta t$=30s gives an interval of 1 day), and a set $\underline{VP} = \{v_1, …, v_o\}$ of VP. The BGP update streams harvested from RIPE are processed to identify 1.) the set of prefixes $Q = \{q_1, …, q_p\}$ for which at least one update is observed at some $v \in \underline{VP}$ during $T \cdot \Delta t$; and 2.) the set of associated AS $A = \{a_1, …, a_m\}$ (i.e., the domains for the $q \in Q$). The update tensor $\underline{X}$ can then be assembled by defining tensor element $x_{ijk}$ to be the number of updates associated with prefixes in AS $a_i$ observed during sampling period j by VP $v_k$. The VP "slices" of $\underline{X}$ are therefore matrices $X_v = [x_1^v \; x_2^v \ldots x_m^v]^T \in \Re^{m \times T}$, where $x_i^v \in \Re^T$ is the update stream involving AS $a_i$ seen at VP v during the observation interval.

Discovering large Internet disruptions involves identifying, for each event, its time of occurrence, affected AS, and observing VP. This discovery is performed by expressing the update tensor $\underline{X}$ as a sum of rank-one tensors via PARAFAC decomposition [25] and then approximating $\underline{X}$ by retaining only the first few terms in the rank-one tensor series. The approximation, denoted $\underline{X}_{approx}$, is a low rank tensor which accounts for most of the variation in the update streams and thus retains any large events. Consequently, examining $\underline{X}_{approx}$ enables large events to be distinguished from the background "noise" associated with normal Internet rerouting activity. Moreover, because $\underline{X}_{approx}$ is expressed in terms of bases for the vector spaces of AS profiles, update streams, and VP profiles, the components of $\underline{X}_{approx}$ allow direct identification of the AS, times, and VP that "participate" in the detected (large) events (see Figure 2).

Consider an Internet disruption r discovered through tensor-based analysis and the set of AS-VP pairs $(AS_r, VP_r)$ associated with this event. In order to locate the network components causing the disruption, we interpret these AS-VP pairs as sets of "end-points" associated with the event and conduct a kind of graph-based "tomographic analysis" [27]. More specifically, for a detected event r, we 1.) label the sets $(AS_r, VP_r)$ on the AS graph $G_{AS}$; 2.) for each $v \in V_{AS}$, compute the *betweenness* [29] of v with respect to the sets $AS_r$ and $VP_r$ (i.e., the betweenness based on all pairs of vertices $(AS_r, VP_r)$); and 3.) arrange the AS in decreasing order of vertex betweenness. The list of AS generated by this procedure forms the (ordered) set of candidate event-causing AS.

This case study illustrates one approach to security-relevant discovery of network dynamics events and offers a principled, automated method for detecting and locating large Internet disruptions. The tensor and graph analysis algorithms are efficient enough to permit implementation at Internet scale in near real-time. Useful insights are generated throughout the analytic process, allowing enhanced utility and flexibility. Additional details regarding algorithm development, implementation, and testing are given in [27]. It is worth mentioning that the proposed tensor-based discovery technique has been successfully applied to several real world events, including the Internet impact of the series of Taiwan earthquakes that occurred in December 2006.

## IV. Situational Awareness

In this section we first define the class of situational awareness (SA) questions of interest, then summarize the proposed Web-based approach to obtaining SA, and finally briefly illustrate our methodology with a case study involving the estimation of public sentiment via analysis of social media content.

### A. Proposed Approach

Many security informatics questions are, at their core, SA problems. For instance, we often wish to know a population's sentiment concerning a situation or how an adversary is framing an issue, as such information helps us to interpret and extrapolate the actions and behaviors of individuals and groups [30]. It is also of interest to identify influential members of a group and authoritative sources of information in a domain, and to recognize collaborating individuals and groups. Additionally, as clandestine behavior is a central concern, it is useful to have capabilities for detecting and defeating deception in its various forms. A common theme in all of these SA problems is the need to make accurate assessments in situations with limited opportunities for observation. Consequently, forming inferences from observables which are incomplete and only indirectly-related to the phenomenon of interest is a key challenge.

For Web-based analysis it is assumed the observables from which SA inferences are to be made are Web objects, and furthermore that these objects possess textual, relational, and/or temporal features. Indeed, a distinguishing aspect of the proposed SA "toolkit" is that it includes methods for analyzing each of these data classes individually and also for integrating and exploiting them in combination.

As an example of a technique for leveraging information present in a single type of data, consider the problem of automatically detecting *framing rhetoric* in the vast volume of Web content generated each day. In [31] we present a simple and effective solution to this problem which consists of three steps: 1.) use the CACTUS system [18] to continually monitor a (potentially large) region of the Web of interest (e.g., a collections of blogs and forums, either specified in advance or discovered as in Section III); 2.) represent the collected content using a bag-of-words model; 3.) identify framing rhetoric (e.g., call-to-action frames) using a naïve Bayes classifier trained on a corpus of exemplar 'framing' and 'non-framing' documents. It is shown in [31] that good classification accuracy (>90%) can be obtained in this way, in part because the producers of framing content in a given domain typically adopt a distinctive and recognizable lexicon in their rhetoric.

Analyzing relational data alone can also yield useful results. We demonstrate in [31] that analyzing hyperlink topology is an effective way to identify the ideology of a given website. More specifically, assume that the SA task of interest is to label websites as belonging to one of two ideological positions, say liberal or conservative in U.S. politics, perhaps so that these sets of pages can be monitored to maintain awareness of the current interests of each camp. Suppose that two sets of exemplar websites are assembled, one for each ideological position, and denote these sets A and B (constructing such sets is typically straightforward to accomplish). Then a simple technique for determining the ideological position of a new website N is to collect together the websites which hyperlink to N and determine whether these sites hyperlink predominantly to the sites in A or to those in B; N is then classified accordingly. This approach is found to accurately distinguish liberal and conservative political sites [31] and is currently being tested in other domains.

Combining content and topology can permit more sophisticated analysis. It is demonstrated in [31] through examination of the Enron email corpus [32] that deceptive users can be reliably identified using a naïve Bayes classifier which takes as inputs both content and topology features. In particular, we find that certain patterns of behavior (e.g., suppression of first-person pronouns in message content, use of 'bcc' for certain addressees) are indicators of deceptive communication [31].

Introducing temporal information allows additional problem domains to be addressed. As a simple example, analyzing time series of documents, for instance blog posts or discussion threads, can reveal trends with important security implications [e.g. 6,17,20,31,33]. Interestingly, we have found that combining topological and temporal data associated with email communication makes it possible to accurately identify collaborating groups of individuals [31]. The underlying observation is simple: a strong indicator that an exchange of communications reflects a genuine collaboration, rather than say a more formal organization-sanctioned affiliation, is a tendency for collaborators to communicate in "bursts" of messages followed by periods of quiet. This pattern can be efficiently recognized by considering time series of who-emails-whom relationships.

Integrating and analyzing all three data types – textual, relational, and temporal – enables even more to be done. For example, consider the task of determining whether a given scientific organization is conducting weapons-related research, perhaps in addition to its legitimate work. It is demonstrated in [20] that one useful approach to this problem is to combine and analyze data on the content of the organization's publications, the co-authorship patterns reflected in these publications, and the way in which publication content and authorship evolves over time. The efficacy of this approach is illustrated in [20] through a study of the Institute of Applied Microbiology in Obolensk, Russia. This assessment indicates that the institute was engaged in bioweapons research during the period 1980-1992, a finding which matches reports by institute defectors (see, e.g., [20,34] for additional details).

Another application which is amenable to analysis that employs all three data types is the identification of influential users in a particular social media domain. In order to make progress it is, of course, necessary to specify what it means for a user to be "influential"; here we adopt the definition that influentials are much more likely than typical users to have their content "go viral", that is, be widely repeated and referenced. Moreover, we require that a putative attribute of influential users have predictive power, so that users which possess this attribute are significantly more likely to produce viral content than are typical users. Recently it has been shown that predictive indicators of influence for bloggers include past influence and blog-graph position [6,35], which implies that determining if a given blogger is influential requires data on blog content, blog graph topology, and the temporal order of posts.

## B. Case Study: Sentiment Analysis

In this case study we consider a central problems in Web-based situational awareness: deciding whether a given social media document expresses positive or negative opinion toward a particular topic. The informal nature of social media content poses a challenge for language-based sentiment analysis. While learning-based methods can provide good performance in settings like this [36], obtaining the required labeled data, such as a collection of "exemplar" blog posts of known sentiment polarity, is an expensive and time-consuming undertaking.

We present in [33] two new methods for inferring sentiment orientation in social media which address these challenges. In each method, documents are represented as bag-of-words feature vectors $x \in \Re^{|V|}$, where the entries of x are the frequencies with which the words in the vocabulary V appear in the document. The goal is to learn a vector $c \in \Re^{|V|}$ such that the classifier orient = $\text{sign}(c^T x)$ accurately estimates the sentiment orientation of document x, returning +1 (−1) for documents expressing positive (negative) sentiment about the topic of interest. Each of the algorithms derived in [33] models the problem data as a bipartite graph $G_b$ of documents/words; the adjacency matrix A for graph $G_b$ is

$$A = \begin{bmatrix} 0 & X \\ X^T & 0 \end{bmatrix} \quad (1)$$

where the matrix $X \in \Re^{n \times |V|}$ is constructed by stacking the document vectors as rows, and each '0' is a matrix of zeros.

The first algorithm adopts a *semi-supervised* approach to sentiment classification, combining knowledge of the sentiment polarity for a few documents and a small lexicon of words with information present in a corpus of unlabeled documents (such unlabeled data are readily obtained online). Assume the initial problem data consists of a corpus of n documents, of which $n_l \ll n$ are labeled, and a modest lexicon $V_l$ of sentiment-laden words, and suppose this label information is encoded as vectors $d \in \Re^{n_l}$ and $w \in \Re^{|V|}$, respectively. Let $d_{est} \in \Re^n$ be the vector of estimated sentiment orientations for the documents in the corpus, and define the "augmented" classifier $c_{aug} = [d_{est}^T \; c^T]^T \in \Re^{n+|V|}$ which estimates the polarity of both documents and words (note that the term $c_{aug}$ is introduced for notational convenience and only the vector c is employed for classification).

We wish to learn an augmented classifier $c_{aug}$ with the following three properties: 1.) if a document is labeled, then the corresponding entry of $d_{est}$ should be close to this ±1 label; 2.) if a word appears in the sentiment lexicon, then the corresponding entry of c should be close to this ±1 sentiment polarity; and 3.) if there is an edge $X_{ij}$ of $G_b$ that connects a document i and a word j and $X_{ij}$ possesses significant weight, then the estimated polarities of i and j should be similar. These objectives can be encoded as a minimization problem:

$$\min_{c_{aug}} \; c_{aug}^T L c_{aug} + \beta_1 \sum_{i=1}^{n_l}(d_{est,i} - d_i)^2 + \beta_2 \sum_{i=1}^{|V_l|}(c_i - w_i)^2 \quad (2)$$

where $L = D - A$ is the graph Laplacian matrix for $G_b$, with D the diagonal degree matrix for A (i.e., $D_{ii} = \Sigma_j A_{ij}$), and $\beta_1, \beta_2$ are nonnegative constants. Minimizing (2) enforces the three properties we seek for $c_{aug}$, with the second and third terms penalizing "errors" in the first two properties. To see that the first term enforces the third property, observe that this expression is a sum of components of the form $X_{ij}(d_{est,i} - c_j)^2$. It is shown in [33] that the $c_{aug}$ (and thus c) which minimizes (2) can be efficiently computed by solving a sparse set of linear algebraic equations (e.g., via the Conjugate Gradient method). Consequently, the semi-supervised approach is implementable with large-scale problems.

The second algorithm proposed in [33] is a novel *transfer learning* method which permits knowledge present in data that has been previously labeled in a related domain to be transferred to a new domain. Assume the initial problem data consists of a corpus of $n = n_T + n_S$ documents, where $n_T$ is the (small) number of labeled documents available for the *target* domain of interest and $n_S \gg n_T$ is the number of labeled documents from some related *source* domain. In addition, suppose that a modest lexicon $V_l$ of sentiment-laden words is known. Let this label data be encoded as vectors $d_T \in \Re^{n_T}$, $d_S \in \Re^{n_S}$, and $w \in \Re^{|V|}$, respectively. Denote by $d_{T,est} \in \Re^{n_T}$, $d_{S,est} \in \Re^{n_S}$, and $c \in \Re^{|V|}$ the vectors of estimated sentiment orientations for the target and source documents and the words, and define the augmented classifier as $c_{aug} = [d_{S,est}^T \; d_{T,est}^T \; c^T]^T \in \Re^{n+|V|}$.

We wish to learn an augmented classifier $c_{aug}$ with the following properties: 1.) if a source document is labeled, then the corresponding entry of $d_{S,est}$ should be close to this ±1 label; 2.) if a target document is labeled, then the corresponding entry of $d_{T,est}$ should be close to this ±1 label, and the information encoded in $d_T$ should be emphasized relative to that in the labels $d_S$; 3.) if a word appears in the sentiment lexicon, then the corresponding entry of c should be close to this ±1 sentiment polarity; and 4.) if there is an edge $X_{ij}$ of $G_b$ that connects a document i and a word j and $X_{ij}$ possesses significant weight, then the estimated polarities of i and j should be similar. The four objectives listed above may be realized by solving the following minimization problem:

$$\min_{c_{aug}} \; c_{aug}^T L c_{aug} + \beta_1 \|d_{S,est} - k_S d_S\|^2 + \beta_2 \|d_{T,est} - k_T d_T\|^2 + \beta_3 \|c - w\|^2 \quad (3)$$

where $\beta_1, \beta_2, \beta_3, k_S$, and $k_T$ are constants and all other terms are as defined above. Analogously to the semi-supervised case, minimizing (3) enforces the properties we seek for $c_{aug}$. The $c_{aug}$ (and thus c) that minimizes (3) can be efficiently computed by solving a sparse set of linear equations.

In [33] the utility of the two proposed learning algorithms is demonstrated by showing they outperform several standard methods for the task of sentiment estimation in two commonly-used test corpora: a set of online movie reviews and a collection of consumer product reviews. Additionally, we illustrate the potential of the methods for security informatics applications through a case study in which sentiment analysis of Arabic, Indonesian, and Danish language blogs is used to estimate regional public opinion regarding the recent revolution in Egypt.

## V. PREDICTIVE ANALYSIS

Prediction problems, and in particular predictions about social systems, are a central concern in security informatics. In this section we consider predictive analysis of two classes of social processes: group dynamics on networks and adversarial coevolution. We begin by offering an approach to prediction which overcomes key deficiencies of standard methods, then briefly review several results obtained by using this approach in Web-based applications, and finally summarize a study involving the prediction of "viral" information diffusion events.

### A. Proposed Approach

Vast resources are devoted to the task of predicting the outcomes of social dynamics, but the quality of such predictions is often quite poor. We show in [21,22] that a fundamental difficulty with standard prediction methods is that they do not rigorously assess problem *predictability*. As a result, these methods frequently fail to identify measurables that possess predictive power, and are therefore unsuccessful because they base their forecasts upon data which are not in fact predictive.

We have recently developed a predictive analysis methodology which addresses this deficiency. Given a question of interest, the proposed analytic procedure begins by determining whether the associated phenomenon is predictable and identifying those measurables (if any) that have predictive power. With these issues resolved it then becomes fairly straightforward to pursue prediction questions, for instance involving early warning analysis or the forecasting of likely outcomes. In Web implementations, the goal is usually to form predictions about "real-world" events based upon measures of online activity.

As a first class of systems of interest, consider social processes in which individuals pay attention to the behavior of others, for instance to obtain benefits of coordinated action or infer otherwise inaccessible information. One of our most interesting findings in this setting is that those system features commonly believed to be predictive are actually not, while other features, related to a system's early dynamics, enable useful prediction. More precisely, it is demonstrated in [21] that the social influences at work in collective dynamics processes act to *reduce* the predictive power of the "intrinsic" attributes of the process (e.g., option quality in social choice situations) and to *increase* the extent to which outcomes can be predicted based upon very early process dynamics [17,20-22]. For example, it is shown in [21,17,6] that in social diffusion (e.g., of information, behavior, or disease), predictability depends crucially upon the topology of the underlying social network, and in particular on this network's community [24] and core-periphery [37] structures (see Figure 3 and [21,17,6]).

We have applied these findings to a broad range of prediction problems. For instance, the papers [21,22] investigate several cultural markets, including those for music and movies. In each market it is found that no measure of a product's intrinsic quality possesses much predictive power. However, the manner in which very early "buzz" about a product diffuses relative to the community and core/periphery structures of the relevant online networks is strongly predictive of the ultimate success (e.g., market share, revenue) of a given product.

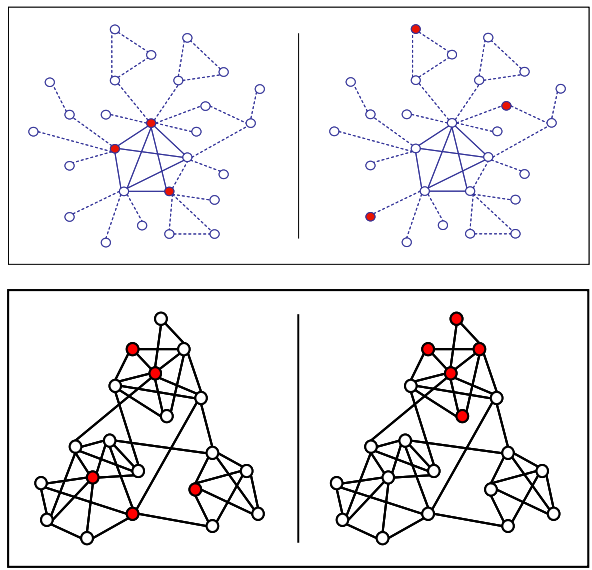

**Figure 3.** Early diffusion within a network's core and dispersed across its communities are predictive. The cartoon at top illustrates the basic idea for the core: diffusion initiated with a set of (red) "seeds" is much more likely to propagate widely if these seeds reside within the core (left) rather than at the periphery (right). Cartoon at bottom depicts the basic idea for communities: diffusion initiated with a set of (red) seeds is much more likely to propagate widely if these seeds are dispersed across multiple communities (left) rather than concentrated within one community (right).

We have obtained similar results in investigations of more security-relevant questions. The emergence and growth of social and political movements, such as the Swedish Social Democratic Party and East German revolution, are examined in [21,17], and it is shown that accurate predictive analysis can be performed by focusing on measures of social network dynamics (as opposed to, say, intrinsics like demographics or movement appeal). Analogous conclusions hold for the dynamics of technology adoption [20,31] and global epidemics [21,17]. This approach also reveals that reliable early warning for real-world mobilization and protest events can be obtained by monitoring diffusion of social media discussions, especially the way discussions propagate over online graph communities [17].

A second class of social dynamics of substantial importance in security informatics are those that define adversarial coevolution, for example in cyber security, terrorism, border security, counterintelligence, and counterproliferation. There is great interest to develop proactive approaches to competing in these "arms races", in which future adversary strategies are anticipated and these insights are incorporated into defense designs. We have recently initiated research in this area, with a focus on cyber security. Perhaps our most interesting finding to date is a mathematically rigorous demonstration that previous attacker actions and defender responses can indeed generate predictive information about future attacker behavior, but only in certain circumstances (see [38] for an example in which such dynamics fails to produce exploitable predictive structures).

As an illustrative example of what can be done in settings in which coevolution leads to predictability, consider the problem of protecting computer networks against intrusions and other attacks. In [39] we leverage the coevolutionary relationship between attackers and defenders to derive two new methods for proactive network defense. The first method is a bipartite graph-based transfer learning algorithm that enables information concerning previous attacks to be "transferred" for application against novel attacks, thereby substantially increasing the rate with which defense systems can successfully respond to new attacks. The second approach involves exploiting basic threat information (e.g., from cyber security analysts) to generate "synthetic" attack data for use in training defense systems, resulting in network defenses that are effective against both current and (near) future attacks.

*B. Case Study: Emerging Topics Analysis*

Discovering and characterizing emerging topics and trends through analysis of Web data is of great interest in both security-related and non-security domains. Recently the paper [40] proposed that monitoring social media to spot emerging *memes* – distinctive phrases which act as "tracers" for discrete cultural units – can enable early discovery of new topics and trends, and presented an effective, scalable algorithm for detecting memes. However, a challenge with this method is the fact that the vast majority of online memes attract very little attention, and in most applications we are interested in those memes, and the underlying topics, that reach a nontrivial fraction of a population of interest. Practical emerging topic discovery therefore requires the ability to identify those memes which will go on to attract substantial attention, and to do so early in the meme lifecycle, as this capability would enable the early detection of *significant* emerging topics and trends.

In [6] we present a new predictive methodology which exploits information about online network topology and dynamics to accurately forecast which memes will propagate widely, appearing in hundreds or thousands of blog posts and/or news items, and which will not. The particular network features used by the prediction algorithm are those identified as likely to be predictive of meme success by our recently developed predictability analysis procedure [21]. Interestingly, the metrics nominated by this theoretical analysis turn out to be fairly subtle measures of the network dynamics associated with early meme diffusion. Meme prediction is accomplished with a machine learning algorithm that, based upon very early network dynamics, is able to accurately distinguish memes which will ultimately diffuse widely from those that will not. Here we briefly indicate the efficacy of the algorithm proposed in [6] through a case study involving detection and analysis of memes associated with an emerging cyber threat.

On 27 December 2008 Israel initiated an air strike against the Gaza Strip, triggering outrage in significant portions of the Muslim world. At the time, national security analysts and officials expressed interest in discovering and characterizing social media discussions which called for retaliations against Israel, particularly those involving "influential" individuals and groups.

To enable a preliminary investigation along these lines, we wrote a Perl program implementing the meme detection algorithm presented in [40] and employed this program to identify memes associated with 'Israel' and 'attack' in a broad range of languages, including Arabic, English, Farsi, French, German, Indonesian, and Turkish. This data identification and collection effort returned a large number of memes, a few of which were classified by our prediction algorithm as being likely to attract significant attention. Intriguingly, most of the memes predicted to "go viral" involved *cyber* attacks on Israel, for instance exhorting Muslim hackers to attack Israeli government and commercial web sites. Example memes detected and analyzed in this way include 'harrased [sic] by Denmark' and '2485 (web)sites' (the latter meme is a reference to a much repeated claim by one hacker group that it had defaced 2485 Israeli websites).

A focused manual examination of the URLs which mention these (and other) memes produced some interesting findings. For example, this analysis led to the discovery of Arabic and Indonesian websites which contain downloadable hacking tools and detailed instructions on the use of these tools. Additionally, we discovered blog posts calling for particular classes of cyber attacks; attacks belonging to these classes were subsequently reported to have taken place by various news agencies.

ACKNOWLEDGEMENTS

This work was supported by the U.S. Department of Defense, the U.S. Department of Homeland Security, and the Laboratory Directed Research and Development Program at Sandia National Laboratories. We gratefully acknowledge many helpful discussions on aspects of this work with Max Planck of the New Mexico Institute of Mining and Technology, Darryl Ackley of the State of New Mexico, and Jim Gosler, Curtis Johnson, and Zach Benz of Sandia National Laboratories.